\documentstyle[12pt,epsf]{article}
\epsfverbosetrue
\def\thebibliography#1{\par\section*{References\@mkboth
  {REFERENCES}{REFERENCES}}\list
  {[\arabic{enumi}]}{\settowidth\labelwidth{[#1]}\leftmargin\labelwidth
    \advance\leftmargin\labelsep
    \usecounter{enumi}}
    \def\newblock{\hskip .11em plus .33em minus -.07em}
    \sloppy
    \sfcode`\.=1000\relax}

\def\citenum#1{{\def\@cite##1##2{##1}\cite{#1}}}
\def\citea#1{\@cite{#1}{}}

\newcount\@tempcntc
\def\@citex[#1]#2{\if@filesw\immediate\write\@auxout{\string
\citation{#2}}\fi
  \@tempcnta\z@\@tempcntb\m@ne\def\@citea{}\@cite{\@for\@citeb:=#2\do
    {\@ifundefined
       {b@\@citeb}{\@citeo\@tempcntb\m@ne\@citea\def\@citea{,}{\bf ?}
\@warning
       {Citation `\@citeb' on page \thepage \space undefined}}%
    {\setbox\z@\hbox{\global\@tempcntc0\csname b@\@citeb\endcsname
\relax}%
     \ifnum\@tempcntc=\z@ \@citeo\@tempcntb\m@ne
       \@citea\def\@citea{,}\hbox{\csname b@\@citeb\endcsname}%
     \else
      \advance\@tempcntb\@ne
      \ifnum\@tempcntb=\@tempcntc
      \else\advance\@tempcntb\m@ne\@citeo
      \@tempcnta\@tempcntc\@tempcntb\@tempcntc\fi\fi}}\@citeo}{#1}}
\def\@citeo{\ifnum\@tempcnta>\@tempcntb\else\@citea\def\@citea{,}%
  \ifnum\@tempcnta=\@tempcntb\the\@tempcnta\else
   {\advance\@tempcnta\@ne\ifnum\@tempcnta=\@tempcntb \else \def
\@citea{--}\fi
    \advance\@tempcnta\m@ne\the\@tempcnta\@citea\the\@tempcntb}\fi\fi}

\def\@maketitle{\newpage
 \null
 \vskip 1em                 
 \begin{flushright}
  {\normalsize \@date}      
 \end{flushright}
 \vskip 2em                 
 \begin{center}
  {\Large\bf \@title \par}  
  \vskip 1.5em              
  {\large                   
   \lineskip .5em           
   \begin{tabular}[t]{c}\@author
   \end{tabular}\par}
\end{center}
 \par
 \vskip 1.5em}                
\def\abstract{\if@twocolumn
\section*{Abstract}
\else \normalsize
\fi}
\def\endabstract{\if@twocolumn\fi\par\clearpage}

\def\section{\@startsection {section}{1}{\z@}{3.5ex plus 1ex minus
    .2ex}{2.3ex plus .2ex}{\Large\bf}}
\def\subsection{\@startsection{subsection}{2}{\z@}{3.25ex plus 1ex minus
   .2ex}{1.5ex plus .2ex}{\large\bf}}
\def\subsubsection{\@startsection{subsubsection}{3}{\z@}{3.25ex plus
1ex minus .2ex}{1.5ex plus .2ex}{\normalsize\sl}}

\def\rules{\@startsection{rules}{5}{\z@}{-3.25ex plus
-1ex minus -.2ex}{0.1ex plus .2ex}{\normalsize\sf}}

\def\appendix{\par\clearpage
  \setcounter{section}{0}
  \setcounter{subsection}{0}
  \@addtoreset{equation}{section}
  \def\@sectname{Appendix~}
  \def\theequation{\thesection\arabic{equation}}
  \def\thesection{\Alph{section}}}

\def\thefigures#1{\par\clearpage\section*{Figures\@mkboth
  {FIGURES}{FIGURES}}\list
  {Fig.~\arabic{enumi}.}{\labelwidth\parindent\advance
\labelwidth -\labelsep
      \leftmargin\parindent\usecounter{enumi}}}

\def\thetables#1{\par\clearpage\section*{Tables\@mkboth
  {TABLES}{TABLES}}\list
  {Table~\Roman{enumi}.}{\labelwidth-\labelsep
      \leftmargin0pt\usecounter{enumi}}}

\def\@sect#1#2#3#4#5#6[#7]#8{\ifnum #2>\c@secnumdepth
     \def\@svsec{}\else
     \refstepcounter{#1}\edef\@svsec{\@sectname\csname the#1\endcsname
.\hskip 1em }\fi
     \@tempskipa #5\relax
      \ifdim \@tempskipa>\z@
        \begingroup #6\relax
          \@hangfrom{\hskip #3\relax\@svsec}{\interlinepenalty \@M #8\par}
        \endgroup
       \csname #1mark\endcsname{#7}\addcontentsline
         {toc}{#1}{\ifnum #2>\c@secnumdepth \else
                      \protect\numberline{\csname the#1\endcsname}\fi
                    #7}\else
        \def\@svse=chd{#6\hskip #3\@svsec #8\csname #1mark\endcsname
                      {#7}\addcontentsline
                           {toc}{#1}{\ifnum #2>\c@secnumdepth \else
                             \protect\numberline{\csname the#1\endcsname}\fi
                       #7}}\fi
     \@xsect{#5}}

\def\@sectname{}

\long\def\@caption#1[#2]#3{\par\addcontentsline{\csname
  ext@#1\endcsname}{#1}{\protect\numberline{\csname
  the#1\endcsname}{\ignorespaces #2}}\par
  \begingroup
    \@parboxrestore
    \small
    \@makecaption{\csname fnum@#1\endcsname}{\ignorespaces #3}\par
  \endgroup}

\def\d{\mathop{\rm d}}

\def\beq{\begin{equation}}
\def\eeq{\end{equation}}
\def\bea{\begin{eqnarray}}
\def\eea{\end{eqnarray}}

\def\bbbone{{\mathchoice {\rm 1\mskip-4mu l} {\rm 1\mskip-4mu l}
{\rm 1\mskip-4.5mu l} {\rm 1\mskip-5mu l}}}
\def\bbbz{{\mathchoice {\hbox{$\sf\textstyle Z\kern-0.4em Z$}}
{\hbox{$\sf\textstyle Z\kern-0.4em Z$}}
{\hbox{$\sf\scriptstyle Z\kern-0.3em Z$}}
{\hbox{$\sf\scriptscriptstyle Z\kern-0.2em Z$}}}}
\def\plb#1#2#3{    {\it Phys. Lett. }{\underbar{B#1}} (19#2) #3}

\relax
\begin{document}
\newcommand{\s}{\\ \vspace*{-2mm} }
\newcommand{\nn}{\noindent}
\newcommand{\nnb}{\nonumber}
\newcommand{\lf}{\left(}
\newcommand{\rg}{\right)}
\newcommand{\as}{\alpha_{\rm S}}
\newcommand{\al}{\alpha}
\newcommand{\p}{\prime}
\newcommand{\asy}{{\rm as}}
\newcommand{\QPM}{{\rm QPM}}
\newcommand{\VMD}{{\rm VMD}}
\newcommand{\pT}{p_{\perp}}
\newcommand{\Dg}{\Delta^{\gamma}}
\newcommand{\Dgp}{\Delta^{\!\gamma}}
\newcommand{\Sg}{\Sigma^{\gamma}}
\newcommand{\Gig}{G^{\gamma}}
\newcommand{\Ftwo}{F_2^{\gamma}}
\newcommand{\Flong}{F_L^{\gamma}}
\newcommand{\La}{\Lambda}
\newcommand{\rar}{\rightarrow}
\def\d{\mathop{\rm d}}
\def\bbbm{{\rm I\!M}}
\def\bbbone{{\mathchoice {\rm 1\mskip-4mu l} {\rm 1\mskip-4mu l}
{\rm 1\mskip-4.5mu l} {\rm 1\mskip-5mu l}}}
%
\centerline{ {\bf  RESOLVED \ VIRTUAL \ PHOTON \ CONTRIBUTIONS}}

\centerline{  {\bf  TO \ INELASTIC \ $ep$ \ SCATTERING.    }
\footnote{To be published in the Proceedings of {\it XV Meeting
          on Elementary Particle Physics}, Kazimierz, Poland, May 1992.}}

\vspace*{1.0cm}
\centerline{\bf Francesca M. Borzumati}
\vspace*{0.2cm}
\centerline{{\it II.\ Institut f\"ur Theoretische Physik
\footnote{Supported~by~the~Bundesministerium~f\"ur~Forschung~und
           Technologie,~05~5HH~91P(8),~Bonn,~FRG.   } }}
\centerline{{\it Universit\"at Hamburg, 2000 Hamburg 50,
 Germany}}
\vspace*{1.0mm}
\begin{center}
\parbox{13.5cm}
{\begin{center} ABSTRACT \end{center}
\vspace*{-1mm}
{\small
\nn
The problem of dealing with the resolved photon contribution
to $P^2$-integrated $ep$ cross sections is discussed. Suitable
approximations have to be found, since no parametrization of the photon
structure function in the full range of virtuality of the
exchanged photon exists.
                                }}
\end{center}

The existence of two substantially different components in the
photon contribution to $ep$ scattering has been widely
discussed in the literature \cite{FAKE}.
One of the two components, the so called {\it direct} one, relies
on the description of the photon as an elementary object, with a
point-like coupling to quarks and leptons. At a fixed order in
$\as$, this component is obtained by evaluating all the
Feynmann diagrams which contribute to the particular $ep$ cross
section of interest.

The second component, the so-called
{\it resolved} one, is based on the assumption of a more
complicated structure of the photon. This picture is corroborated
by experimental evidence of the behaviour of the photon as a
meson. The resolved photon component is given in terms of the
{\it photon structure function} (PHSF). In a physical gauge, the PHSF
can be represented by the sum of all the so-called ladder diagrams
starting with a hadronic input or a point-like one, i.e. the
elementary vertex $\gamma qq$. Since this vertex is typically used
in the evaluation of the direct component of the photon contribution,
it is clear that problems of double counting may arise in the
calculation of $ep$ cross sections.
Surprisingly enough, the existence of this problem was recognized
only relatively recently\cite{FAKE}.

Even more recently, several simple but important observations were
brought to attention in ref.~\cite{MESCH}. Some of them are discussed
in this talk. No matter what type of procedure is adopted to get rid
of the double counting \cite{FAKE}, it is not completely clear how to
deal with the resolved photon contribution, or with the PHSF, when
calculating cross sections integrated over the full range of virtuality
($P^2$) of the exchanged photon.
Typically, values of $P^2$ from $\sim 10^5\,$Gev$^2$ down to almost
zero are involved in $ep$ scatterings at HERA.
In principle one can numerically deal with the region
of real PHSF, $P^2\ll \La^2\ll Q^2$, where $Q^2$ plays the
role of the typical scale of the process, and the region of
virtual PHSF, $\La^2\ll P^2\ll Q^2$. Several parametrizations
exist, in fact, for the real PHSF and known is also the analytical
expression of the virtual PHSF in moment space. In practice, though,
the solution of the virtual PHSF in $x$-space is also not available,
as yet.

Two more regions of $P^2$, besides the ones already mentioned, are
involved in the calculation of photon mediated $ep$ processes:
$P^2 \sim \La^2 \ll Q^2$ and $\La^2 \ll P^2 \sim Q^2$. It is

\vspace{0.5cm}
\epsfysize=20cm
\epsfbox{singl80.eps}
\vspace{0.5cm}
\begin{small}
\nn
{{\bf Fig.\,1}
  \ Singlet distribution as function of $P^2$
  at $Q^2 = 80\,$GeV$^2$ for the different moments $n=2,4,6$. The right
  solid lines denote the virtual singlet solution. The straight solid
  lines are the real singlet distribution function obtained from the
  LAC1 input \cite{LAC}. The dotted lines are the QPM result
  with the logarithmic term continued to
  $\log(Q^2/(P^2 +\La^2))$. From ref.~\cite{MESCH}.  }
\end{small}

\vspace{0.5cm}
\epsfysize=11.0cm
\epsfbox{gluon40.eps}
\vspace{0.5cm}
\begin{small}
\nn
{{\bf Fig.\,2}
  \ The gluon distribution in the virtual
  photon as a function of the photon virtuality at
  $Q^2 = 40\,$GeV$^2$ ($n_f = 4$ and $\La = 0.2\,$GeV). The solid,
  dashed and long-dashed line, refer to $n=2,4$ and $6$, respectively.
  The values of the three moments of the LAC1
  parametrization \cite{LAC} of the
  gluon distribution in the real photon are $17.1$ ($n=2$), $0.18$
  ($n=4$), and $0.045$ ($n=6$). From ref. \cite{MESCH}. }
\end{small}
\vspace*{0.5cm}

\nn intuitive that for high enough values of $P^2$, i.e.\ for
$P^2 \sim Q^2$, the
quark parton model result should give a valid description of the PHSF.
In contrast, the continuation of the real and
virtual solution over the region $P^2 = {\cal O}(\La^2)$ is
non-trivial. For practical purposes, though, one may think of
stretching the two solutions obtained in the case of real and virtual
PHSF over this region.

That this works relatively well for quark distribution functions
is shown in Fig.~1, where the $P^2$ dependence of the
known solutions for the singlet distribution
$\Sg (n,Q^2\!\!,P^2)$ is shown for different moments $n$. I remind
that, at the leading order in $\as$, $\Sg(n,Q^2\!\!,P^2)$ is
expressed in terms of the distributions of the quark and
anti-quark in the photon as
\beq
          \Sg (n,Q^2\!\!,P^2)   =
        \sum_i \left[ q^\gamma_i(n,Q^2\!\!,P^2)+
           \overline{q}^\gamma_i(n,Q^2\!\!,P^2)\right]\ .
\label{singl}
\eeq
As can be seen, the virtual solution has been extrapolated in
Fig.~1 down to $\La^2$ and the real one, as obtained
from the LAC1 parametrization \cite{LAC}, has been extended up to
$(10^{-1}\,\mbox{GeV})^2$. It can also be observed that the QPM
result for $\Sg (n,Q^2\!\!,P^2)$, with the logarithmic term
continued to $\log{(Q^2/(P^2\!+\!\La^2))}$, exhibits a very slow
onset of a $P^2$ dependence and a smooth behaviour over the region
$P^2 \sim \La^2$. Given the features shown in this figure, the
matching of the real and virtual distribution functions does not seem
so unreasonable. The precise value of $P^2$ where the matching can be
performed, called $P_{\rm cut}^2$ in ref.~\cite{MESCH}, depends on
the particular real PHSF parametrization used, on $n$ and $Q^2$.
The matching point increases with increasing $n$ and decreases
with increasing $Q^2$. One may push this simplification one step
further and approximate the virtual
singlet distribution $\Sg (n,Q^2\!\!,P^2)$ above $P_{\rm cut}^2$ with
the QPM solution. Of course, slightly overestimated values for
the total and differential cross sections should be expected.

The $P^2$ dependence of the gluon distribution function, shown in
Fig.~2, has a much more dramatic behaviour, dropping to
zero quite fast. It is clear from this figure how the extrapolation
of the gluon distribution in a real photon throughout the full range
of $P^2$ is not adequate. Since the $x$-space expression for the
virtual gluon distribution is still lacking, a better but quite
crude solution, at the moment, seems to simply drop the gluon
distribution above the value $P_{\rm cut}^2$ chosen for the quark
distributions. This type of strategy is again suggested by the QPM
approximation. Underestimated cross sections, with discontinuous
$P^2$ distributions, are then obtained.

To conclude, few words should be said about the sensitivity of
$P^2$-integrated $ep$ cross sections to the type of approximations
mentioned here. This sensitivity is obviously process-dependent. To
begin with, the uncertainty depends on the size of the
``uncertainty interval'' for $P_{\rm cut}^2$ relative to the total
$P^2$ range. Moreover, quark-initiated processes can be calculated more
reliably than gluon-initiated ones because of the softer $P^2$
dependence of quark distribution functions. An important role is
also played by the direct photon contribution to these cross sections.
The larger is this
contribution, the less is the cross section affected by these
approximations (and the procedure used to eliminate the double counting
problem mentioned at the beginning). An important parameter in this
respect is the value of the typical scale of the process $Q^2$. This
scale determines the effective upper and lower limits of the $P^2$
integration (see \cite{MESCH}), the size of the direct contribution
with respect to the resolved photon ones, and influences also
the uncertainty in the choice of $P_{\rm cut}^2$. Finally, as shown in
\cite{MESCH}, high values of $Q^2$ make the approximation of the
virtual photon distributions with the ones obtained in the QPM worse
than in the case of small scales $Q^2$.

Obviously, the uncertainty in the evaluation of photon-mediated
$ep$ cross sections could be reduced if a parametrization of the PHSF
in the full range of virtuality needed at HERA and future colliders
would be available. More work in this direction is needed.
\begin{small}

\end{small}


\begin{thebibliography}{99}
%
\bibitem{FAKE}
  Due to lack of space, for this and other points mentioned in this
  talk, the redear is referred to \cite{MESCH,TEUPI} and the
  references therein listed.

\bibitem{MESCH}
 F.M. Borzumati and G.A. Schuler, DESY preprint, DESY 92-078,
  (1992), to be published in {\it Zeit. f\"ur Physik}.

\bibitem{TEUPI}
 F.M. Borzumati, DESY preprint, DESY 92-095 (1992), to be
  published in the proceedings of the Zeuthen Workshop on
  Elementary Particle Theory, Teupitz, Germany, April 6-10, 1992.

\bibitem{LAC}
  H. Abramowicz, K. Charchula and A. Levy, \plb{269}{91}{458}.

\end{thebibliography}
\end{document}